\documentclass[submission,copyright,creativecommons]{eptcs}
\providecommand{\event}{MARS 2024} 

\usepackage{iftex}
\usepackage{csquotes}

\usepackage{tikz}
\usetikzlibrary{arrows}
\usetikzlibrary{arrows.meta}
\usetikzlibrary{automata}
\usetikzlibrary{calc}
\usetikzlibrary{chains}
\usetikzlibrary{decorations.pathmorphing}
\usetikzlibrary{decorations.pathreplacing}
\usetikzlibrary{matrix}
\usetikzlibrary{patterns}
\usetikzlibrary{petri}
\usetikzlibrary{positioning}
\usetikzlibrary{scopes}
\usetikzlibrary{shadows}
\usetikzlibrary{shapes}
\usepgflibrary{shapes.multipart}
\usetikzlibrary{shapes.symbols}
\usetikzlibrary{through}
\definecolor{actionrow}{gray}{0.9}
\definecolor{disabled}{gray}{0.8}
\definecolor{error}{rgb}{1,0,0}
\tikzstyle{every edge}=[draw,semithick,>=latex,shorten >=1pt,shorten <=1pt]
\tikzstyle{every initial by arrow}=[initial distance=7mm,initial text=]
\tikzstyle{every place}=[draw=blue,fill=blue!20,minimum size=6mm]
\tikzstyle{every transition}=[draw=red,fill=red!20,minimum width=10mm,minimum height=0mm]
\tikzstyle{error}=[color=error]
\tikzset{%
	dblarw/.style n args={2}{
		-{Circle[scale=0.6]},
		line width=1.5pt,
		draw=#2,  
		color=#2, 
		opacity=1,
		postaction={
			draw=#1,
			color=#1,
			line width=1pt,
			shorten >=0.5pt,
			shorten <=0.3pt,
		},
		dblarw/.default={black}{white},
		dblarw/.initial={black}{white},
	}
}

\usepackage{amsmath}
\usepackage{placeins}
\usepackage{adjustbox}
\usepackage{subfig}
\usepackage{siunitx}
\usepackage{listings}
\usepackage[regular,ligatures=TeX,scale=.9]{sourcecodepro}
\renewcommand*\copyright{{%
		\fontfamily{lmr}\selectfont\textcopyright}}

\ifpdf
  \usepackage{underscore}         
  \usepackage[T1]{fontenc}        
\else
  \usepackage{breakurl}           
\fi

\newcommand{\grant}{This study was partially funded by DFG SCHU 2479, and DFG SCHL 1844/6-1.}
\newcommand{\figshare}{\url{https://doi.org/10.6084/m9.figshare.25061336}}

\title{Sliced Online Model Checking for Optimizing the\\Beam Scheduling Problem in Robotic Radiation Therapy}
\author{Lars Beckers$^1$
\qquad
Stefan Gerlach$^2$
\qquad
Ole Lübke$^1$
\and
Alexander Schlaefer$^2$
\qquad
Sibylle Schupp$^1$
\email{\{lars.beckers, stefan.gerlach, ole.luebke, schlaefer, schupp\}@tuhh.de}
\institute{$^1$Institute for Software Systems \qquad\qquad $^2$Institute of Medical Technology and Intelligent Systems\\
Hamburg University of Technology, Hamburg, Germany\thanks{\grant}}
}
\def\titlerunning{Sliced Online Model Checking for Optimizing Beam Scheduling}
\def\authorrunning{L. Beckers et al.}

\newcommand{\uppaal}{\mbox{\textsc{Uppaal}}}
\begin{document}
\maketitle

\begin{abstract}
In robotic radiation therapy, high-energy photon beams from different directions are directed at a target within the patient. Target motion can be tracked by robotic ultrasound and then compensated by synchronous beam motion. However, moving the beams may result in beams passing through the ultrasound transducer or the robot carrying it. While this can be avoided by pausing the beam delivery, the treatment time would increase.
Typically, the beams are delivered in an order which minimizes the robot motion and thereby the overall treatment time. However, this order can be changed, i.e., instead of pausing beams, other feasible beam could be delivered.

We address this problem of dynamically ordering the beams by applying a model checking paradigm to select feasible beams. Since breathing patterns are complex and change rapidly, any offline model would be too imprecise. Thus, model checking must be conducted online, predicting the patient's current breathing pattern for a short amount of time and checking which beams can be delivered safely. Monitoring the treatment delivery online provides the option to reschedule beams dynamically in order to avoid pausing and hence to reduce treatment time.

While human breathing patterns are complex and may change rapidly, we need a model which can be verified quickly and use approximation by a superposition of sine curves. Further, we simplify the 3D breathing motion into separate 1D models. We compensate the simplification by adding noise inside the model itself. In turn, we synchronize between the multiple models representing the different spatial directions, the treatment simulation, and corresponding verification queries.

Our preliminary results show a \SIrange{16.02}{37.21}{\percent} mean improvement on the idle time compared to a static beam schedule, depending on an additional safety margin. Note that an additional safety margin around the ultrasound robot can decrease idle times but also compromises plan quality by limiting the range of available beam directions. In contrast, the approach using online model checking maintains the plan quality. Further, we compare to a naive machine learning approach that does not achieve its goals while being harder to reason about.

\end{abstract}

\section{Introduction}
Radiation therapy presents a widely used option for cancer treatment. Typically, beams from a range of different directions are used to deliver a therapeutically effective dose to the tumor while maintaining a tolerable dose for all other tissues. The latter is particularly important for critical structures where radiation damage would result in severe side effects. While different treatment systems have been proposed, one interesting approach is robotic radiosurgery where a robotic arm carries the beam source, allowing for a very large solid angle of possible beam directions. The optimal choice of beams results from careful treatment planning\cite{Schlaefer2008a}, which accounts for the beams’ attenuation when passing through the patient.

\begin{figure}[tb]
	\centering
	\begin{minipage}{.5\textwidth}
		\centering
		\includegraphics[width=.8\textwidth,clip,trim={0 6cm 0 0}]{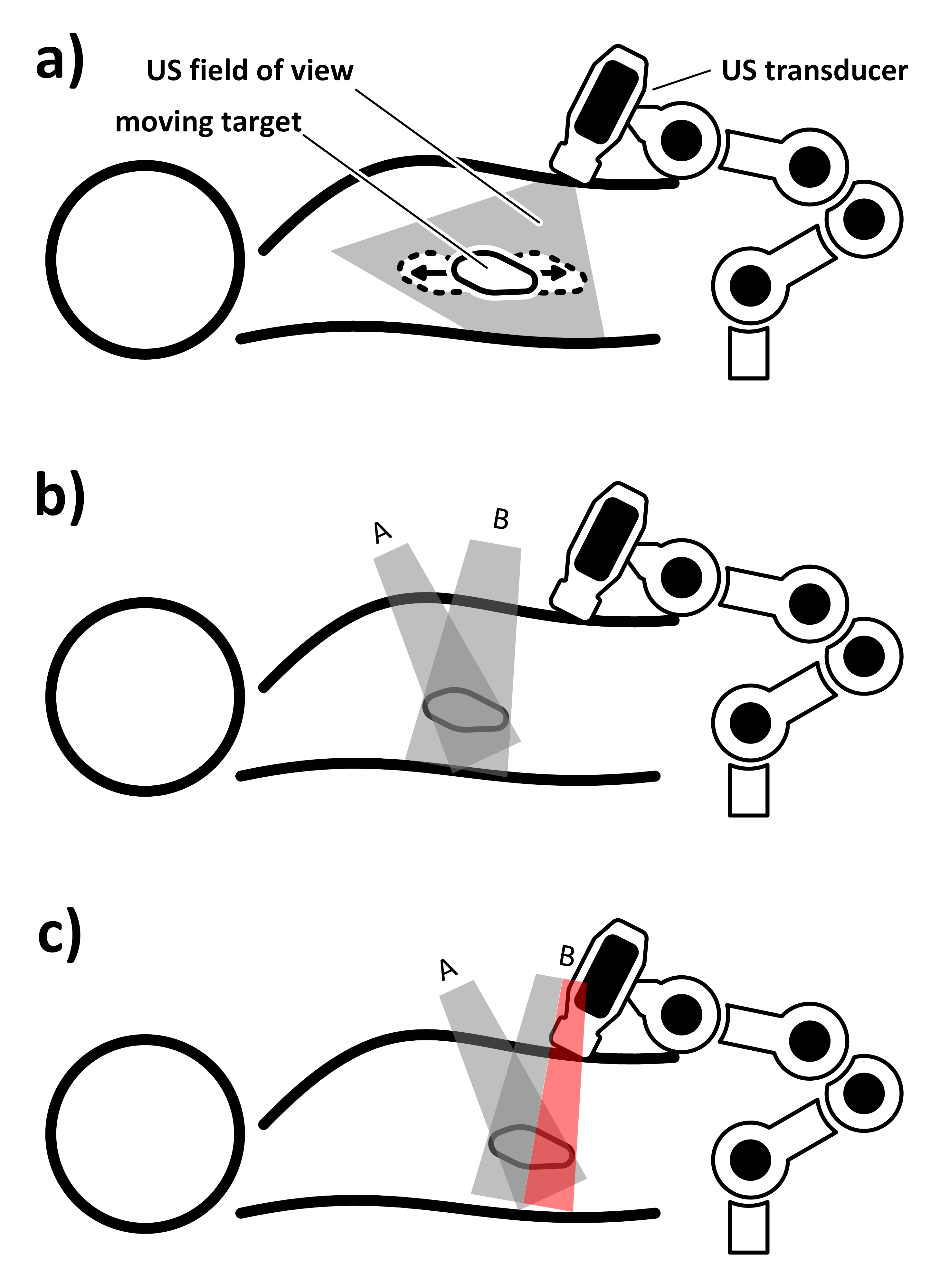}
	\end{minipage}%
	\begin{minipage}{.5\textwidth}
		\centering
		\includegraphics[width=.8\textwidth,clip,trim={0 0 0 12cm}]{sketch_of_the_problem}
	\end{minipage}
	\caption{Illustration of the situation we consider. Figure\,a) shows the the general setup with the patient and the robot mounted ultrasound (US) transducer, as well as the target which moves along the patient's superior-inferior axis. Figure\,b) depicts a position of the target where two beams, A and B, are both feasible and can be delivered. Figure\,c) shows another target position for which beam B would be partially blocked and therefore infeasible. Note, that the range of motion of the target may vary and for some intervals during the overall treatment, beam B may also be feasible throughout the full breathing cycle.}
	\label{fig:problem-illustration}
	\vspace{-1em}
\end{figure}

One of the particular advantages of robotic beam delivery is the ability to adjust the position and orientation of the beam source quickly. This is desirable, as some tumors are subject to substantial motion, especially due to respiration and in regions close to the diaphragm. Different respiratory motion prediction and control strategies have been studied to realize real-time motion compensated treatments, which are now routinely used with the robotic \textit{CyberKnife} (Accuray, USA)\cite{Ernst2013}. However, a key aspect of motion compensation is the detection of the tumor’s actual motion. Typically, this is achieved by a combination of X-ray imaging and external marker tracking. However, recently the use of robotic ultrasound has been considered, which would allow fast, continuous and non-ionizing imaging of the tumor’s motion\cite{Gerlach2017a,Gerlach2022,Gerlach2023}. This advantage comes at the cost of integrating the ultrasound transducer and the robot carrying it into the overall system setup. Particularly, safe beam delivery must be guaranteed at all times, i.e., as Figure~\ref{fig:problem-illustration} shows, the treatment beams must not pass through ultrasound transducer and robot, as this would compromise the dose estimate.

At planning time, the overall collision free delivery of all beams can be optimized, i.e., a generalized traveling salesman problem (GTSP) is solved to determine the best sequence of beams and to obtain the coordinated motion of both robots\cite{Schlueter2019a}. However, during treatment the motion trajectories of the tumor vary, e.g., as patients may change from chest to abdominal breathing. A straightforward approach to maintain the planned dose is to pause the treatment beam whenever the motion causes the current beam to collide with the ultrasound transducer or robot. Given that this prolongs the overall treatment time and that other beams may be feasible, it would be desirable to predict the feasibility of collision free beam delivery given the current tumor motion pattern and to select beams which can be safely delivered without interruption.

In this setting, we propose studying model checking as a means to realize safe and effective treatments. Instead of using a fixed order on the beams, we consider modeling the current respiratory motion and to verify for each beam whether it is feasible, i.e., can be delivered.  While model checking verifies properties---regarding safety, in particular---of modeled systems, in case of a breathing patient it is impossible to provide a full model of all possible future behavior. Thus, if we want to verify if a beam is feasible during a treatment session, we need to reduce the scope of this verification, both in terms of detail represented by the model and in terms of the covered time span, down to the next few seconds. Still, we need to regularly verify whether beams are feasible based on current motion data to make up for inaccurancies and to cover the whole treatment session. This approach of repeated verification with current data is called \textit{Online Model Checking}.

While the basic breathing motion is similar among patients, the specifics of their situation are different e.g., whether their inhalation is short or deep, fast or slow, how much it varies over time, when exactly, how grave the differences between those patterns are. Overall, a model will necessarily be inaccurate and the online checking approach is used to limit this effect. The predicted patient's motion---as the foundation to strategically select beams---needs to be formally verified to ensure safety, whether it stays within the safe region. Moreover, at any time it would be important to consider a number of beams in order to determine some feasible beam to be delivered next, because verification of a single beam may not be sufficient to get an optimal or even viable beam. At the same time it is possible that multiple beams satisfy the safety requirements, requiring a decision as part of an optimization problem.

We want to satisfy above requirements within a software system that improves the overall treatment delivery. For this software, its timeliness is important, since the online verification is necessarily limited in its future scope and we need to offer new beams in a timely manner. Instead of terminating when the time allotted for the session is insufficient, it will use the available time more efficiently and thus allow for more completed treatments.
Based on these requirements and existing tools, our work resulted in the following contributions:

\vspace{-6pt}
\begin{itemize}\setlength\itemsep{0pt}
	\item An approach for a verified schedule of beams in radiation therapy. This approach is based on an existing online model checking environment for respiratory motion. We contribute the statistical verification of dynamically selected beam candidates in real time.
	\item A representation of 3-dimensional respiratory motion as product of three 1-dimensional model slices to limit the model's complexity. This requires synchronization of verfication queries and results within the real time environment.
	\item An implementation in Rust, using the \uppaal{} model checker for verification. Further, the implementation of experiments including data generation for evaluation of supporting our approach with machine learning methods.
\end{itemize}
\vspace{-6pt}

The result is a dynamic beam scheduling which selects feasible beams and thereby reduces the treatment time. Upcoming beams are selected according to our online model of patient motion data, while the timing requirements of the approach are honored. We can keep the timing requirements by the use of a simple, 1D model that allows us to keep verification times short. In turn, we necessarily increase the complexity of our implementation that needs to synchronize the 3 parallel online models and its beam verification tasks. We propose an architecture that combines three parallel, model-generating processes, the verification processes for beams as required, and the actual client that simulates the therapy.
We reduce idle time where no beam can be delivered due to collisions by \SIrange{16.02}{37.21}{\percent} depending on the amount of additional safety margin around the ultrasound robot.

Furthermore, we discuss an approach to improve the beam selection by using AI models to make predictions on the breathing patterns and beam verification times. This would allow us to make better decisions in our selection and verification. We tried to train a multi-layer perceptron regressor on beam verification duration, as well as a classifier on 1D breathing movement data to discern breathing pattern irregularities into different categories. However, this was not successful to derive meaningful strategies from: The regressor predicted verification durations on the lower end of their range and the classifier did not learn the intended categories, where more difficult patterns are characterized by noisier or rapidly changing movements. Here, an algorithm detecting extreme values may be better suited.

In the following section, we provide details of our beam radiation therapy case study. In section~\ref{sec:omc}, we present the necessary background to our approach. This is followed by sections~\ref{sec:model} and \ref{sec:data}, in which we discuss our models and data, respectively. In section~\ref{sec:app}, we elaborate on the architecture and implementation of our contribution. The experimental results are covered in section~\ref{sec:experiments}. In section~\ref{sec:ai}, we discuss possible machine learning enhancements to our approach. Finally, we close with a summary in section~\ref{sec:summary}.

\section{Ultrasound Guided Motion Compensated Radiation Therapy}\label{sec:therapy}
\begin{figure}
	\centering
	\begin{minipage}{.5\textwidth}
		\centering
		\begin{tikzpicture}
			\node[inner sep=0,outer sep=0] (image) {\includegraphics[width=\textwidth,keepaspectratio]{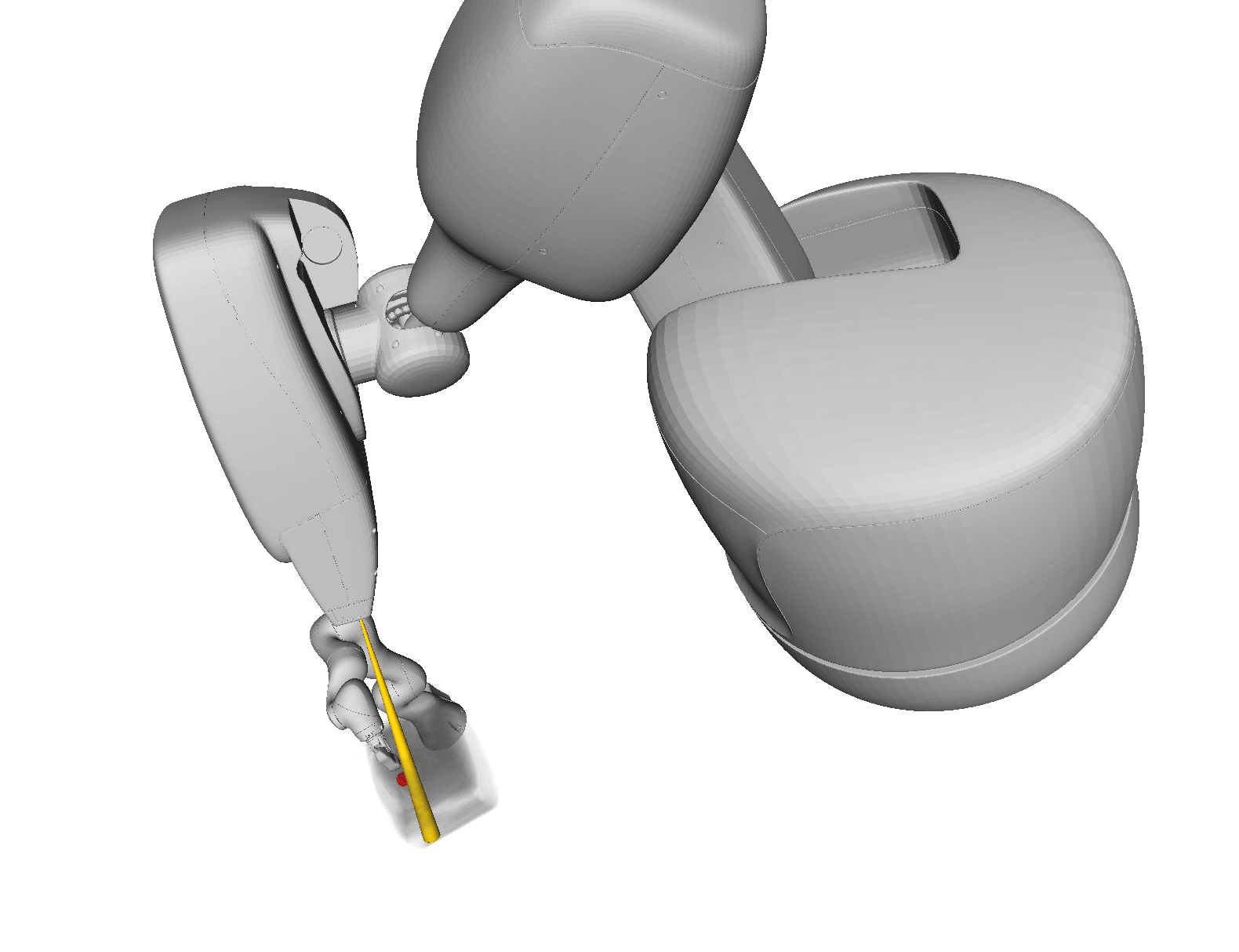}};
			\draw[dblarw={black}{white}] (2.4,2.3) node[above]{\textcolor{black}{treatment robot}} -- (2.2,1);
			\draw[dblarw={black}{white}] (-1.8,-2.5) node[below]{\textcolor{black}{ultrasound robot}} -- (-1.75,-1.4);
			\draw[dblarw={black}{white}] (-1.2,-0.5) node[right]{\textcolor{black}{beam}} -- (-1.63,-1.3);
			\draw[dblarw={black}{white}] (0,-1.9) node[right]{\textcolor{black}{patient}} -- (-1.2,-1.9);
		\end{tikzpicture}
		\caption{Two interacting robots: One robot is used for treatment delivery and the other robot is used for ultrasound image guidance.}
		\label{fig:beam-apparatus}
	\end{minipage}%
	\begin{minipage}{.5\textwidth}
		\centering
		\vspace{-1em}
		\begin{lstlisting}[basicstyle=\ttfamily\scriptsize,numbers=none,frame=none]
			ID,Time[ms],Threshold[mm]
			80731,24281,5.0
			76503,11222,5.0
			75681,13682,7.5
			74528,23749,10.0
			67108,2243,10.0
			79427,7133,12.5
			70571,16927,15.0
			77460,4354,15.0
			70211,16240,15.0
			68851,1488,17.5
			59592,7335,17.5
			74430,14448,17.5
			69894,29738,20.0
			77674,1609,20.0
			78301,16381,22.5
			81561,3116,25.0
			61025,17047,27.5
			71430,1758,31.0
			81038,21519,31.0
		\end{lstlisting}
		\vspace{-1em}
		\caption{Excerpt from a 1D beam list.}
		\label{fig:beam-list}
	\end{minipage}
	\vspace{-1.5em}
\end{figure}

In radiation therapy, multiple beams are directed at a target region within the human body, which is typically referred to as a \textit{planning target volume} (PTV). The cumulative dose delivered by the beams is optimized to reach a therapeutically effective level inside the PTV, while being as low as possible outside and particularly in so-called \textit{organs at risk} (OAR). During plan optimization the beams are weighted to realize a clinically acceptable trade-of between these objectives. For our sample scenario we consider robotic beam delivery with the CyberKnife and robotic ultrasound based on the lightweight and redundant \textit{LBR iiwa med robot} (KUKA, Germany) as illustrated in Figure~\ref{fig:beam-apparatus}.

During a treatment session, a patient inhales and exhales and may exhibit varying breathing patterns, e.g., slower and faster breathing, deep and shallow breathing, abdominal and chest breathing, and a number of potentially sudden changes, e.g., due to coughing. When considering motion compensation, the movement of the PTV is off-set by an equivalent motion of the beams, i.e., the beam carrying robot moves the beams synchronously. Hence, the beams move relative to all static objects, including the ultrasound transducer and the robot carrying it. However, beams need to be switched off whenever they would pass through these objects, as the effective dose would be compromised by the additional attenuation and scattering.

While an effective treatment requires all beams to be delivered, there is typically no requirement to observe a particular order. In fact, the order is often optimized such that the time to visit all beam starting positions is minimized, as the motion time adds to overall treatment time. In case of ultrasound guidance the coordinated motion of both robots can be approached as a GTSP and a time-optimal beam schedule can be computed\cite{Schlueter2019a}. However, as the breathing patterns are not known a priori, it may occur that some beams cannot be delivered without interruption. At the same time, other beams may be feasible. Any beam that satisfies the safety requirements and does not collide with ultrasound transducer or robot is a viable alternative beam to be delivered. Thus, we consider beam scheduling as the optimization problem to minimize the time beams are inactive due to collisions.

Consider, for example, a patient exhibiting an increasingly deeper breathing motion during the treatment session. We can compute the minimum distance between beam and ultrasound transducer and robot at any time in the breathing cycle, which we refer to as threshold. Now, starting the session with beams with large thresholds and unsuccessfully trying to apply beams with small thresholds later will potentially be a worse solution than a randomized ordering. Furthermore, the robot requires a substantial amount of time to change its configuration between beams, particularly when the next beam is far away from the previous beam.

Therefore, our proposal is a dynamic list of beams that is checked for possible beams within the next time slot. While it is preferable to continue along the current time-optimal list, especially finishing a currently running beam, we can substitute waiting times where no beam is applied with other beams that can be delivered. Afterwards, the new time-optimal list of beams can be computed and treatment continued. Thus, safety requirements are upheld while the treatment duration can be reduced.

\vspace{-1em}
\section{Online Model Checking}\label{sec:omc}
In online model checking (OMC), properties of a modeled system are verified regularly in an interval\cite{OMCDependable}. The system processes exhibit uncertainties. Thus, statistical model checking is typically used, which assesses the probability of its queried property over multiple, simulated runs of its system until it has established the necessary confidence. This behaviour can be customized through the configuration of the model checker, but we do not employ this in this paper. Each verification is limited in its temporal scope to reduce the required verification time to fit within the interval. Thus, the model is only valid for a short amount of time before the drift from reality gets too large. Still, the temporal scope needs to be large enough to ensure that in each verification step there is enough time for an emergency stop if verification fails and the sequence of verification intervals overlaps to cover the progressing system entirely.
OMC has already been used in several different applications\cite{OMCIntegration}\cite{OMCAccelerating}\cite{LightweightHybridMC}, including cyber-physical systems and medical settings\cite{CPSHybridMC}\cite{MedicalHybridMC}. However, it had not been used for beam verification.

For proper verification, a model is regularly modified with new data. In reality, this data may be supplied by sensors recording live events. In experimental settings, it can be either recorded data or entirely synthetic. The most common modification is to update and reassign model variables that represent the current state, e.g., the patient's position. Thus, for a continuing online setting, we receive a line of models synthesized from the template, each representing one time step in the verification process.

For our setting, we make use of an existing online model checking software that predicts the respiratory motion of a patient in its primary movement dimension\cite{Rinast2015}. It acts as a frontend to the underlying statistical model checker, \uppaal{}-SMC\cite{uppaal}\cite{uppaal-smc}: It receives data updates to provide these to the model, implements reading in model templates and writing out the modified instances every three seconds for a validity window of six seconds. As it has been used to demonstrate respiratory motion\cite{Antoni2016}, the template model and required data sampling and transformation have also already been implemented. Thus, adaptions to our problem have remained minimal.
The OMC software also verifies that its modified model is an adequate representation according to its input data. It does so by advancing time by one second within the interval, and assessing the probability that the observed patient position is within a bounding box of the expected position. For completeness, we provide the queries and corresponding evaluation model in Appendix~\ref{apx:bbox}, but regard this in the following as a black\,box step, as only its verdict is relevant to us.

\section{Modeling}\label{sec:model}
In the following section, we will provide details on the respiratory motion model and how we use the generated motion models for our beam scheduling problem, so that our new software can assess the deliverability of potential beams w.r.t. the motion.
We approximate respiratory motion as a superposition of sine curves in space, in each of the three dimensions. We present two modeling designs that simplify the representation of respiratory motion and  explain how oversimplification is prevented. Furthermore, we describe the verification of the beams.

\begin{figure}[tb]
	\centering
	\begin{adjustbox}{width=\textwidth}
	\begin{minipage}{.5\textwidth}
		\centering
		\includegraphics[width=\textwidth,clip,trim={0 .5cm 0 .5cm}]{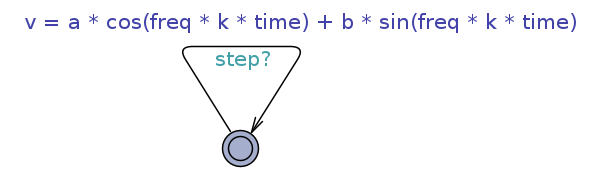}
		\caption*{FSTerm}
		\label{fig:fsterm}
	\end{minipage}%
	\begin{minipage}{.5\textwidth}
		\centering
		\includegraphics[width=.75\textwidth,clip,trim={0 .55cm 0 .42cm}]{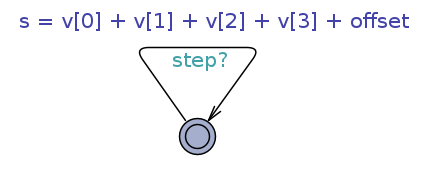}
		\caption*{Summer}
		\label{fig:summer}
	\end{minipage}
\end{adjustbox}
\begin{adjustbox}{width=\textwidth}
	\begin{minipage}{.35\textwidth}
	\centering
	\includegraphics[width=\textwidth,clip,trim={0 .45cm 0 .33cm}]{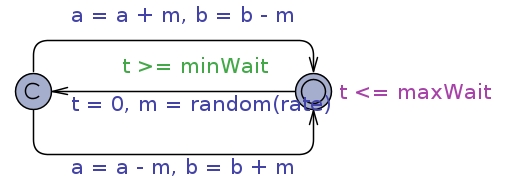}
	\caption*{TermModifier}
	\label{fig:termmodifier}
	\end{minipage}%
	\begin{minipage}{.35\textwidth}
	\centering
	\vspace{-1.4cm}
	\includegraphics[width=.65\textwidth,clip,trim={0 .5cm 0 .4cm}]{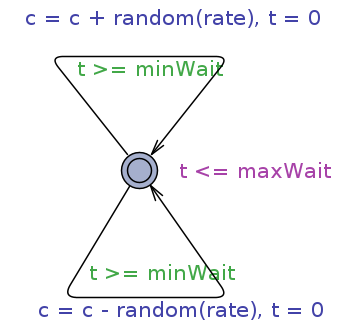}
	\caption*{CoeffModifier}
	\label{fig:coeffmodifier}
	\end{minipage}%
	\begin{minipage}{.35\textwidth}
	\centering
	\vspace{-.1cm}
	\includegraphics[width=\textwidth,clip,trim={0 .6cm 0 .5cm}]{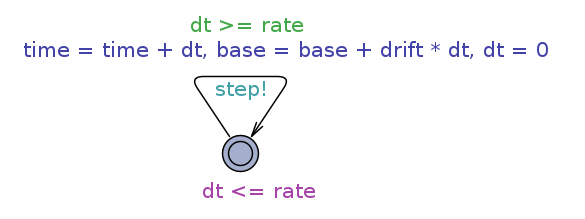}
	\caption*{Timer}
	\label{fig:timer}
	\end{minipage}
\end{adjustbox}
	\caption{The complete 1D model: \uppaal{} automata network template representing the respiratory motion.\cite{Rinast2015} Its declarations are given in Appendix~\ref{apx:declarations}. The continually updated variables are also shown in Figure~\ref{fig:3d-modeling}. Note the time stepping through \uppaal{}'s synchronization messages.}
	\label{fig:respiratory-model}
	\vspace{-1em}
\end{figure}

\vspace{-10pt}
\subsection{1D motion modeling}\label{sec:1dmodeling}
\vspace{-6pt}
The first simplification, introduced in \cite{Antoni2016}, concerns the reduction to a 1D motion representation through modification of the motion model shown in Figure~\ref{fig:respiratory-model}. The represented axis is orthogonal to the patient's body and is the primary movement direction, while the other two dimensions behave similarly but with less extreme motion. The idea is that the orthogonal axis is the most important one. In each step, the OMC software samples new positions, which would be measured by the treatment controller in reality, as it progresses in time, and updates this base model accordingly.

The model is a simplified breathing motion representation based on the following formula:
\vspace{-.5em}
\begin{align*}
	x(t) = d \cdot t + \sum_{k = 1}^{4} c_k \cdot cos(k \cdot f \cdot t) + s_k \cdot sin(k \cdot f \cdot t),~~
	\text{with } t \text{ time},~ d \text{ linear drift},~ f = \frac{2\pi}{period}.
\end{align*}
\vspace{-1em}

\noindent The position samples are Fourier-transformed to obtain the decomposed \textit{s} and \textit{c} terms of a sine and a cosine wave over time at the samples' frequency. Four terms are summed and must later be offset by a base value, which the wave moves around.
In Figure~\ref{fig:respiratory-model}, we already see all necessary components to describe this as an \uppaal{} model. The \textit{Summer} automaton just computes a sum \textit{s} when a new step is due. This sum is assigned to the model's global \textit{result} variable. The sum's first four components are computed by four instances of \textit{FSTerm}, which computes the four combined sine-cosine waves of the formula, where \textit{a} and \textit{b} are the transformed samples, \textit{k} is the ordinal number of the term, \textit{freq} the sample's frequency, and \textit{time} is a stepped view into the clock. Those summed together with the base position, parameterised as \textit{offset} here, is enough to compute the patient's position. The \textit{Timer} automaton is used to drive this computation in steps at a regular interval of $38$ milliseconds, parameterised as \textit{dt}, and moves the base according to its expected \textit{drift}.

However in Figure~\ref{fig:respiratory-model}, we also see two additional automata.
In order to compensate that the real motion does not adhere to a (simplified) mathematical model based on an exact formula, the model includes an \textit{accuracy} parameter that determines how much randomness is applied to the coefficients and terms. Thus, when setting the accuracy value below 100\%, a range of possible motion curves is considered starting from the latest current data point. This is important, because not only the model itself is simplifying, but also the validity period of at most six seconds is already quite long considering that a patient may suddenly start coughing. The accuracy value is set at the start of a session, thus a line of models created for a single patient will have a fixed value.
The \textit{TermModifier} automaton is instantiated for all four indices into the \textit{a} and \textit{b} arrays. The values are modified at random, based on a \textit{rate} derived from the accuracy parameter. Instead of regular time steps, there is simply a minimum and maximum wait time between modifications. Similarly, the \textit{CoeffModifier} is instantiated twice to modify both the \textit{base} and \textit{frequency}, represented as \textit{c} here.

Besides the automaton network, the model declares the variables describing the actual movement. The continually updated, declared variables of such a model are shown in Figure~\ref{fig:3d-modeling}.
The \textit{period} value describes the period of the regular breathing sine curve. The \textit{base} value is the origin of this period. The four cosine and sine terms are composed as arrays \textit{a} and \textit{b}. As mentioned, there is a (small) \textit{drift} value added to the base for every time step. Among other intermediate variables, the computed value describing the position of the patient at each step is saved as \textit{result}. Thus in Section~\ref{sec:beam}, we query this value within the relevant time period with respect to the requested threshold for our verification purposes. We provide the full declarations of the model in Appendix~\ref{apx:declarations}.

\begin{figure}
	\centering
	\vspace{-12pt}
	\begin{adjustbox}{width=.9\textwidth, center}
	\begin{minipage}{.35\textwidth}
		\centering
		\begin{lstlisting}[basicstyle=\ttfamily\footnotesize,numbers=none]
const double period = 5088.0;
const double drift = -0.0;

double base = -3.6508;
double a[4] = { -0.608, 0.205, 0.0744, -0.0764 };
double b[4] = { 2.5745, -0.414, -0.0149, 0.0096 };
		\end{lstlisting}
		\vspace{-12pt}
		\caption*{X axis}
		\label{fig:3dx}
	\end{minipage}%
	\begin{minipage}{.35\textwidth}
		\centering
		\begin{lstlisting}[basicstyle=\ttfamily\footnotesize,numbers=none]
const double period = 5088.0;
const double drift = 0.0;

double base = 1.698;
double a[4] = { 0.2631, -0.0887, -0.0322, 0.0331 };
double b[4] = { -1.1144, 0.1792, 0.0065, -0.0041 };
		\end{lstlisting}
		\vspace{-12pt}
		\caption*{Y axis}
		\label{fig:3dy}
	\end{minipage}%
	\begin{minipage}{.35\textwidth}
		\centering
		\begin{lstlisting}[basicstyle=\ttfamily\footnotesize,numbers=none]
const double period = 5088.0;
const double drift = 0.0;

double base = 1.8164;
double a[4] = { 0.0757, -0.0255, -0.0093, 0.0095 };
double b[4] = { -0.3202, 0.0516, 0.0019, -0.0012 };
		\end{lstlisting}
		\vspace{-12pt}
		\caption*{Z axis}
		\label{fig:3dz}
	\end{minipage}
	\end{adjustbox}
	\vspace{-10pt}
	\caption{Declarations of the models of a single timestep of patient \mbox{DB126-Fx1}. As the position and movement is different, the beam verification queries require different thresholds per dimension.}
	\label{fig:3d-modeling}
	\vspace{-1.6em}
\end{figure}

\vspace{-12pt}
\subsection{3D motion modeling}\label{sec:3dmodeling}
\vspace{-6pt}
While the actual respiratory motion is three-dimensional, it often has a clear principle component, i.e., along the superior-inferior direction. Still, the motion may be non-linear and a refined model would consider all three spatial motion components. The second simplifying model design is to represent 3D motion by a network of three 1D models. Because the other two dimensions can be approximated themselves using the same mathematical formula (see \ref{sec:1dmodeling}), we can reuse the 1D model for each instead of extending it to 3D variables and thusly more complex formulas. This allows us to stick with a relatively simple model that can be verified fast enough for our purposes.

However, using one model per dimension means that we need three parallel OMC processes that each generate new, separate models. Since for any online approach we need timestamped, 3-dimensional position data, this kind of input dataset needs to be retrieved by sensors to be distributed among the three processes. As a consequence, the process' results need to be aligned to make sure all models are instantiated, represent the same time slot and are successfully validated.

The synchronization problem extends further to the beam checks that have to be done on all models separately with separate threshold data: In Figure~\ref{fig:3d-modeling}, we show for a single timestep the variable sets of the three models it is comprised of. Because the motion along the three axes has different positions each, each beam has necessarily six threshold values instead of one, i.e., a set of two for each model.

It is beneficial to coordinate the accuracy parameters for the three movement directions, covering the respective inaccuracies as with the 1D case, but also to control that the different dimensions are not moving completely distinctly as the separate models suggest. In the future, we want to merge different possibilities of 3D beam positions to reduce the number of verification calls.

\vspace{-10pt}
\subsection{Beam verification}\label{sec:beam}
The aim of the treatment is to deliver all beams, which requires that at some point during the treatment all beams have become deliverable, i.e., are not blocked due to the motion. Thus, the core verification question is whether the motion remains within the threshold that is given per beam. Our software uses the results of the verification in \ref{sec:1dmodeling} to determine a list of deliverable beams per time slot among a requested list of beams. Along with the online approach, the requested beams are updated by the treatment software each time slot and the verification and selection of beams is repeated until the treatment is complete.

The beams themselves are not formally modeled, but represented by a query on the motion model that we check per beam. In case of 3D, that leads to three queries per beam, i.e., one per dimension.
The query is created directly from the source data: \texttt{Pr[<= \{scope\}] ([] result <= \{upper\} \&\& result >= \{lower\})}, where \textit{scope} is set to three seconds (see \ref{sec:1dmodeling}), \textit{result} is the patient position in one of the three dimensions (see \ref{sec:3dmodeling}), and \textit{lower} and \textit{upper} are derived from the beam threshold (see Figure~\ref{fig:beam-list} for a symmetric 1D case); \texttt{[]} is the usual \enquote{global} operator in CTL. The statistical model checker then determines the probability with which, within \textit{scope}, the patient's position will invariantly be safe.

We use the verification results of the beams to gather a list of deliverable beams that is provided to the treatment software. Representing a safety requirement, beams may not be considered as deliverable if the threshold of any movement direction is violated according to the verification result.
If the resulting probability is higher than our cutoff, a beam may be delivered and is added to the list. We present results in Section~\ref{sec:experiments} with a cutoff of $0.5$, whereas further experiments suggest a cutoff of $0.91$.
In the case of 3D, the deliverability of a beam then becomes the minimum of each of the results on the three distinct models.
It is possible that a query, or in 3D at least one of three queries of a beam, is not completed within the timeslot. In these cases, the beam is also not considered to be deliverable.
However, we observe these queries to take only milliseconds to complete. Depending on actual timings, requested lists of 250 beams often may be completely checked and answered.

\vspace{-12pt}
\section{Data}\label{sec:data}
\vspace{-4pt}
We evaluate on a small amount of seven sets of real patient motion data (966670 data points) which was previously published\cite{Ernst2013} and extend our dataset with synthetic data (4893007 data points) for our machine learning approach.
In\,this\,section, we report\,on our\,available\,datasets and the limitations of synthetic\,data.

\begin{figure}[tb]
	\centering
	\includegraphics[width=.9\textwidth,clip,trim={2.5cm 2.2cm 5cm 4.2cm}]{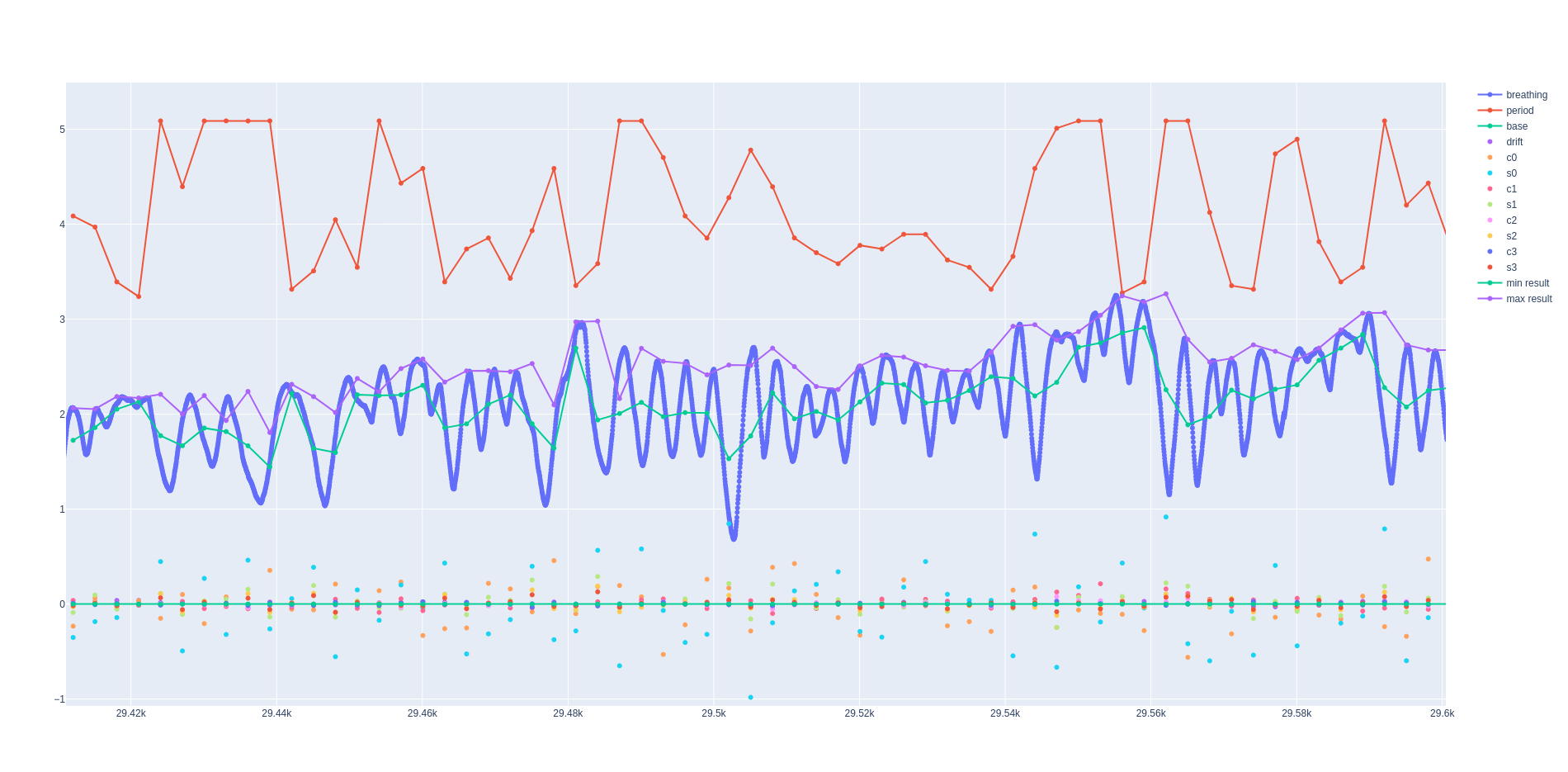}
	\vspace{-10pt}
	\caption{Partial superior-inferior portion of the breathing curve of patient DB126-Fx1 in blue along time on the x axis. Model variable values of period as red line, base as green line through the middle of the blue line, with the sum formula values c0, c1, c2, c3 and s0, s1, s2, s3 dotted. Illustrated with an expectation query of the statistical model checker for the minimal and maximal curve point by the purple line atop the blue line and the green line along 0 on the y axis.}
	\label{fig:data-sample}
	\vspace{-1em}
\end{figure}

The respiratory motion data consists of 3 dimensional data points that are timestamped. Furthermore, we generated 40 synthetic motion curves approximating idealized respiratory motion with relatively simple changes and only in 1D along the superior-inferior axis. For each of these patients we provide\footnote{\figshare} the corresponding line of models generated through the OMC process; for the real patients also in all three dimensions. A sample of this dataset is visualized in Figure~\ref{fig:data-sample}. All lines of models come with their data logs, so that they can be easily visualized and analyzed. Furthermore, we provide beam lists for both 1D and 3D cases.

Generating patient breathing data, however, is not trivial. It is important to consider that not all situations can be well covered as synthetic data is much smoother and easier to align to for the OMC process, and resulting curves are simpler and distinct. Real breathing changes more often, is irregular, and has lots of small changes. This can be seen in Figure~\ref{fig:data-sample}, which is an excerpt of under four minutes. There, the OMC process is constantly changing its predictions and beams differ in their verification results. Thus, it is important to use not only synthetic data to make sure observed behavior is realistic.

Compared to breathing data, beam lists are relatively simple to generate, both for real movement paths and also synthetically. For synthetic beam lists we can use existing beam lists as input to generate new lists. Each beam is distinct from the next, thus we do not need to consider changing movements along a curve. Thus, new lists just need to have the same distribution within the threshold and time values with regards to minimal, maximal, and average values overall and within quantiles.

\vspace{-12pt}
\section{Beam Checking Application}\label{sec:app}
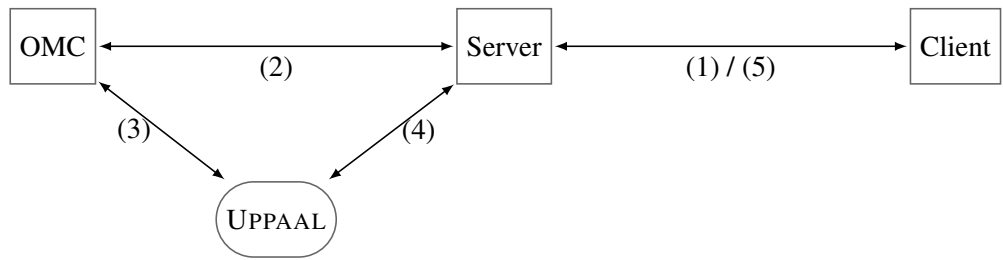
\begin{figure}
	\begin{center}
		\begin{tikzpicture}[on grid, node distance=60mm,draw,semithick,>=latex]
			\node[rectangle, minimum size=1cm, draw=black!60] (omc) {OMC};
			\node[rectangle, minimum size=1cm, draw=black!60, right of=omc] (server) {Server};
			\node[rectangle, minimum size=1cm, draw=black!60, right of=server] (client) {Client};

			\draw (omc) edge[<->] node[below] (between) {(2)} (server);
			\node[rectangle, minimum size=1cm, rounded corners=5mm, draw=black!60, below=20mm of between] (uppaal) {\uppaal};

			\draw (server) edge[<->] node[below] {(1) / (5)} (client);
			\draw (omc) edge[<->] node[left] {(3)} (uppaal);
			\draw (server) edge[<->] node[right] {(4)} (uppaal);
		\end{tikzpicture}
		\vspace{-\baselineskip}
	\end{center}
	\caption{Architecture diagram of the software systems that we employ for our experiments. Each temporal slice consists of:
		(1) Client sends beam list, (2) Server uses Online Model Checking (OMC) to receive models predicting respiratory motion, (3) OMC uses \uppaal{} for model creation and verification, (4) Server starts \uppaal{} processes to verify beams, (5) Server sends list of beam results.
	}
	\label{fig:architecture}
	\vspace{-1.5em}
\end{figure}

\vspace{-6pt}
Our system needs to synchronize between the online model checking (OMC) process that generates the models themselves, the treatment robot, and our own verification of possible beams to apply. The general architecture is explained in this section and visualized in Figure~\ref{fig:architecture}.

The client is externalized as an application that connects over network and sends a current list of beams that are outstanding. In turn, we send a list of checked beams with their success probabilities until the updated beam list is empty. Having the client networked is a proven way for separation of concerns and allows the---possibly pre-existing---client to implement its own, e.g., safety equipment to cover cases such as sudden changes by coughing that cannot fully be anticipated by software. However, this separation induces the need to properly synchronize between these and the state of the session.

The OMC application is run directly by our software and monitored for new model files and its verification of those. It runs self-contained and does not pose requirements on the application. We currently use pre-existing lists of breathing motions that would in reality need live data. The self-validation of the OMC process also means that if verification is not passed for one model, it can recover on one of the next intervals based on updated sensor data, but until it recovers there is no current model against which to verify beams. Thus, we cannot make any recommendation within this gap and it follows that the client would need to pause the treatment, failing on the safe side.

Whenever we receive a new model from the OMC process, we execute a new set of beam verification queries, which check per beam the probability that it never violates its threshold within the timescope.
We align the OMC time slots with our own verification calls, allowing a three second time slot for our own tasks and thusly verifing beams regularly, similar to the OMC process itself. However, the results of this checking process is not a new model but a recommendation for or against those beams. If, e.g., a beam failed to verify within the time slot, either by the property itself or by the end of the verification interval, it is not part of the recommended set of viable beams corresponding to this time slot.

In order to gain as much information as possible within this interval, we execute multiple verification queries in parallel.
Still, it is likely that we may not have enough time to verify all remaining beams, so we prefer certain beams.
First of all, we prioritize the currently running beam if there is one, but also all beams that have already been started. This reduces overhead of beam changes on the treatment robot and increases application quality, because the dosages are affected by application times and cannot precisely be controlled by starts and stops. Second, we select well applicable beams by offering a mixture of short and long applied beams, but also consider their thresholds. Beams with high thresholds, but shorter running times are clearly easier to apply even if breathing is not calm, slow, or steady. Thus for future experiments, we intend to implement a feedback loop that further prioritizes beams according to the actual current breathing situation rather than static properties.

As introduced, our architecture can be characterized as loosely coupled as we kept three separated systems in our architecture. Despite the synchronization issues this architecture necessitates, the separated concerns helped with the implementation of the software.
By using the pre-existing OMC process, we only needed to introduce minimal changes, i.e., updating the code base to run within a more modern environment, including adding Java synchronization primitives and exporting the actual model files.
Similarly, the client is separated from our application. Communication is realized over standard input/output streams of the process in case of the OMC process or over a TCP socket in case of the client. Instead of constraining us to a single programming language, we have specified the file formats of the transmitted data. The OMC application transmits simple log messages and \uppaal{} models in their already specified file format, while the client sends and receives CSV tables with specified attributes.

The server application software is written in Rust. The beam verification queries are executed via the \uppaal{} command line interface with the result parsed from its standard output, compared to the OMC software using \uppaal{}'s in-process library API. The approach via external processes was taken because it allows us to terminate verification processes according to our timeouts as required, which is not possible via the library itself. Furthermore, the independent processes do not require synchronized access to a singular verification server instance that the library allows to access, when evaluating multiple different queries as with beams.

For the client, we extend our in-house treatment planning Java application. We determine beam collision with the ultrasound robot by applying a projection based approach using a distance transform\cite{Gerlach2017a}. Here, we can also specify a static safety margin which accounts for expected target motion. Note however, that this safety margin needs to be specified before the treatment and leads to an elimination of additional beam directions. Therefore, that treatment plan quality degrades with increasing safety \mbox{margin.}

We extend our approach to calculate beam thresholds by computing a projection for every beam translation in the discretized motion trace. The treatment simulation starts by computing a beam order which is time optimal if the target is static, i.e., it minimizes the robot motion. Afterwards, the client cycles through the following steps:
The beam list with corresponding thresholds is sent to the server for evaluation whether delivery will be feasible. After receiving the response from the server, the first feasible beam is selected, the robot is moved in position and delivery is started. Before the next evaluation cycle starts, the time optimal beam order is updated with respect to the current position of the robot.
During beam delivery, it is ensured that the beam is collision free and otherwise beam delivery is halted and the cycle is started again.


\vspace{-12pt}
\section{Experiments}\label{sec:experiments}
\vspace{-6pt}
In our evaluation we first show preliminary results of our approach. We evaluate on 5 patient geometries for which the target tumor is located in the liver. Since treatment planning typically involves the selection of a subset of beams from a large set of randomized candidate beams, we repeat each experiment 30 times with different candidate beam sets to obtain statistically meaningful results. We report significance according to the Student's t-test with p-values smaller than \SI{0.01}{}.

We use a motion trace of captured target motion on a previous CyberKnife treatment\cite{Ernst2013}, where the breathing pattern showed a large change over time. While the particular motion pattern is not representative for every patient, it represents a challenging real-world case for ultrasound guided radiation\,therapy.

To evaluate the impact of our online model checking approach, we simulate complete treatments. We compare a delivery of beams where each beam is scheduled either according to the static beam list or according to our OMC approach. When a scheduled beam cannot be delivered due to the target motion, we pause the beam delivery until the beam is not blocked. We sum up this idle time to draw conclusions on the quality of our approach. Additionally, we track how often beam delivery needs to be interrupted.

Additionally, we evaluate whether the pose of the ultrasound robot influences the results, and whether an additional safety margin could be used to reduce the idle time and the occurrences of beam collisions. Here, we apply a safety margin of \SI{5}{\mm} to the ultrasound robot during treatment planning and not consider beams which are too close to the ultrasound robot. Thereby, small motion of the target would not lead to a collision during treatment. However, this additional margin impacts the treatment plan quality since fewer beams are available during the treatment planning process\cite{Gerlach2017a}.

\begin{figure}[tb]
	\vspace{-6pt}
	\subfloat[idle time]{\includegraphics[width=0.45\textwidth]{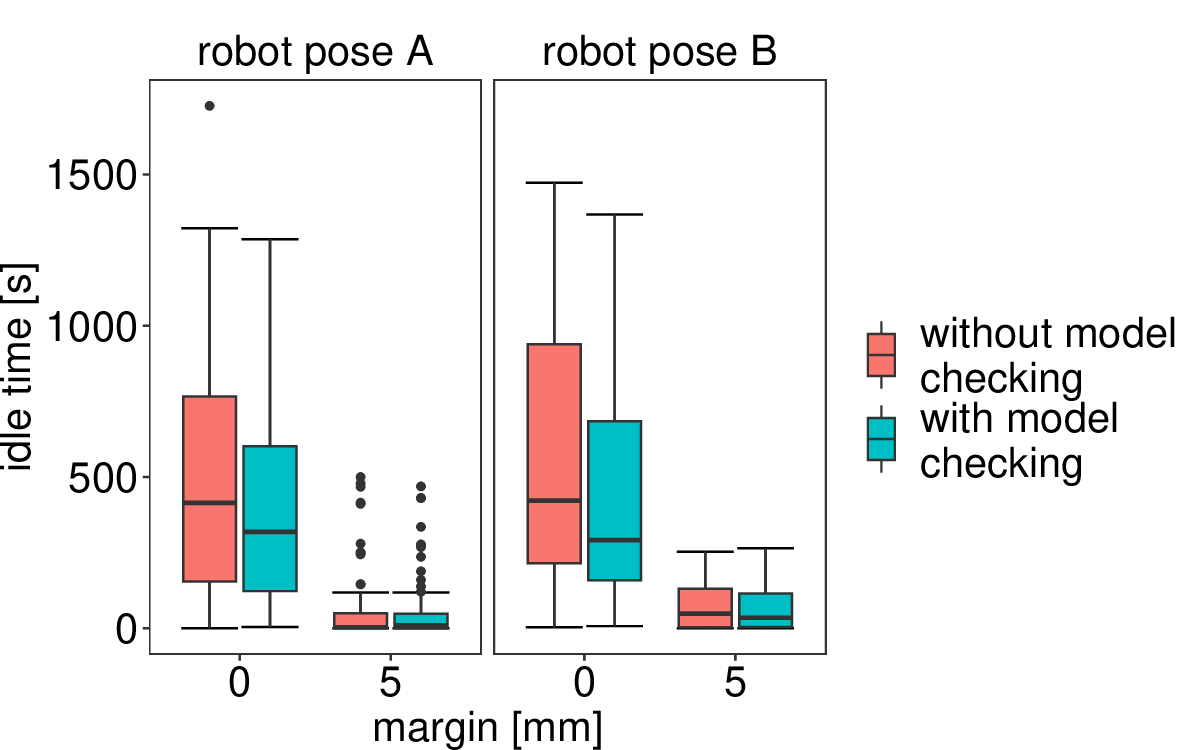}}
	\hfill
	\subfloat[incomplete deliveries]{\includegraphics[width=0.45\textwidth]{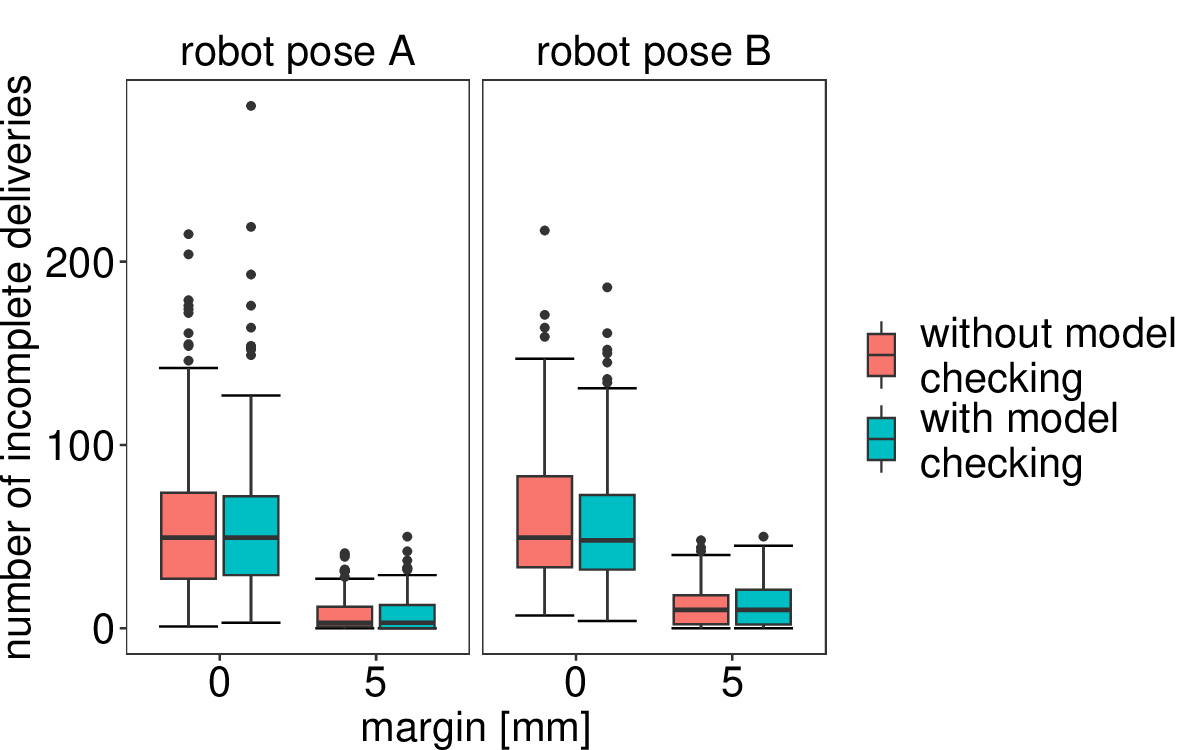}}
	\vspace{-6pt}
	\caption{Summarized idle time (a) and number of incomplete beam deliveries (b) per simulation for the different approaches for 5 patients and motion trace DB126\_Fx1. Experiments are repeated 30 times with different beam sets to increase statistical significance.}
	\label{fig:results:time-impact}
	\vspace{-6pt}
\end{figure}
\begin{figure}[tb]
\begin{minipage}{.45\textwidth}
	\begin{center}
		\includegraphics[width=\textwidth]{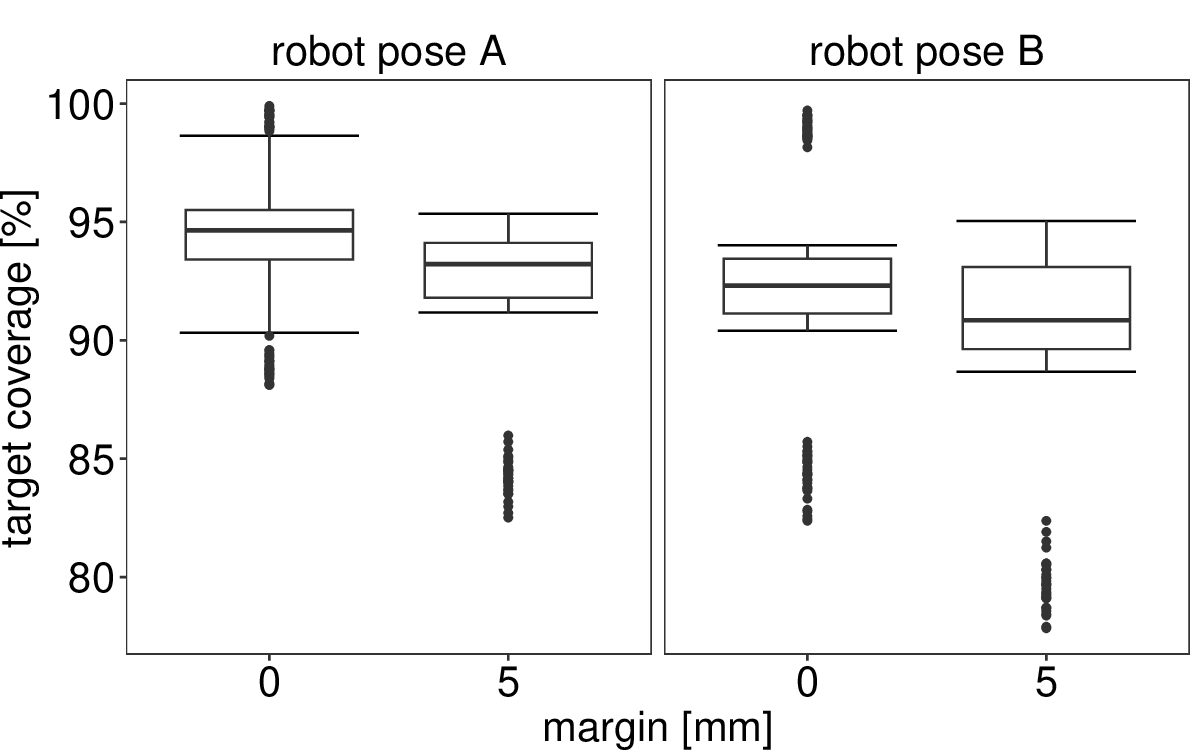}
	\end{center}
	\vspace{-8pt}
	\caption{Example of resulting target coverage for 5 patients. Experiments are repeated 30 times with different beam sets to increase statistical significance.}
	\label{fig:results:target-coverage}
\end{minipage}%
\hfill
\begin{minipage}{.45\textwidth}
	\centering
	\includegraphics[width=.75\textwidth,trim={19cm 2.5cm 23cm 7cm},clip]{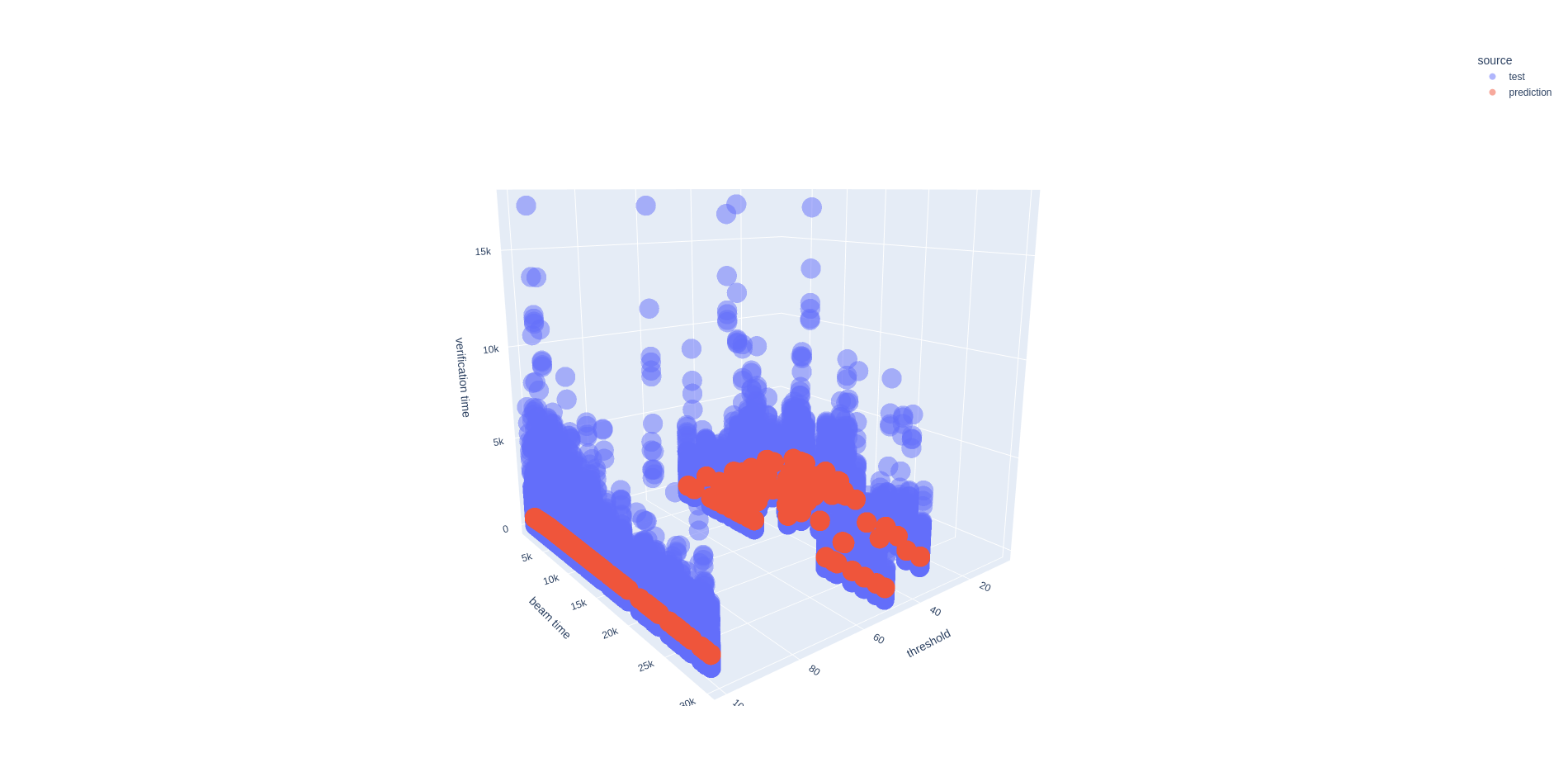}
	\vspace{-6pt}
	\caption{Predictions in red, actual data in blue. Note the difference in transparency: The blue circles are much more varied than the red.}
	\label{fig:ai-regressor}
\end{minipage}
\vspace{-1em}
\end{figure}

Our results in Figure~\ref{fig:results:time-impact}a show that the average idle time decreases from \SIrange{418.44}{304.96}{\second} for no margin and from \SIrange{25.93}{22.35}{\second} for a margin \SI{5}{\mm}. This represents a reduction in idle time of \SI{37.21}{\percent} and \SI{16.02}{\percent}, respectively. While the difference is significant for no margin ($\text{p} < 0.004$) it is not significant when using \SI{5}{\mm} margin ($\text{p} = 0.22$). While the pose of the robot can change the resulting idle time in some cases, e.g., from \SIrange{44.66}{76.95}{\second} for \SI{5}{\mm} margin and no OMC ($\text{p} = 0.001$), the difference is otherwise not significant on the average distribution ($\text{p} > 0.02$).
When evaluating the number of times that a beam could not be delivered shown in Figure~\ref{fig:results:time-impact}b, we do not observe a significant difference between our OMC approach and the static beam ordering ($\text{p} > 0.38$).

As Figures~\ref{fig:results:time-impact} show, increasing the margin around the ultrasound robot also decreases the number of collision events and the resulting idle time. Still, Figure~\ref{fig:results:time-impact}a shows that also when introducing the margin, the idle time is even lower with our OMC approach. Crucially, the target coverage decreases when introducing a margin as Figure~\ref{fig:results:target-coverage} shows. The target coverage refers to the proportion of the target which receives at least the prescribed dose and is an important clinical goal which influences the clinical outcome. Therefore, the safety margin is degrading the treatment quality, while OMC does not negatively influence the delivered dose as per Figure~\ref{fig:results:time-impact}b. Note also that extreme outliers exist for certain patients with substantially worse coverage when applying an increased safety margin.

\section{AI Enhancements}\label{sec:ai}
We tried to extend the information we can gain from our data using machine learning methods. First, we tried an experiment on beam verification time learning. Second, we tried an experiment on breathing pattern classification. Both would allow us to change the prioritization of beams to ensure a more fitting list of verification results or even a larger list of verification results when we exclude beams that are not likely to succeed in verification by failing either time requirements or anticipated threshold violation. Although both experiments failed, we briefly report in this section their respective setting and outcome.
These experiments were implemented in Python to make use of the widely known \textit{scikit-learn}\cite{scikit-learn} library that provides good default implementations of common learning algorithms.

\begin{figure}[tb]
	\centering
	\includegraphics[width=.9\textwidth,trim={2.3cm 2cm 5cm 12cm},clip]{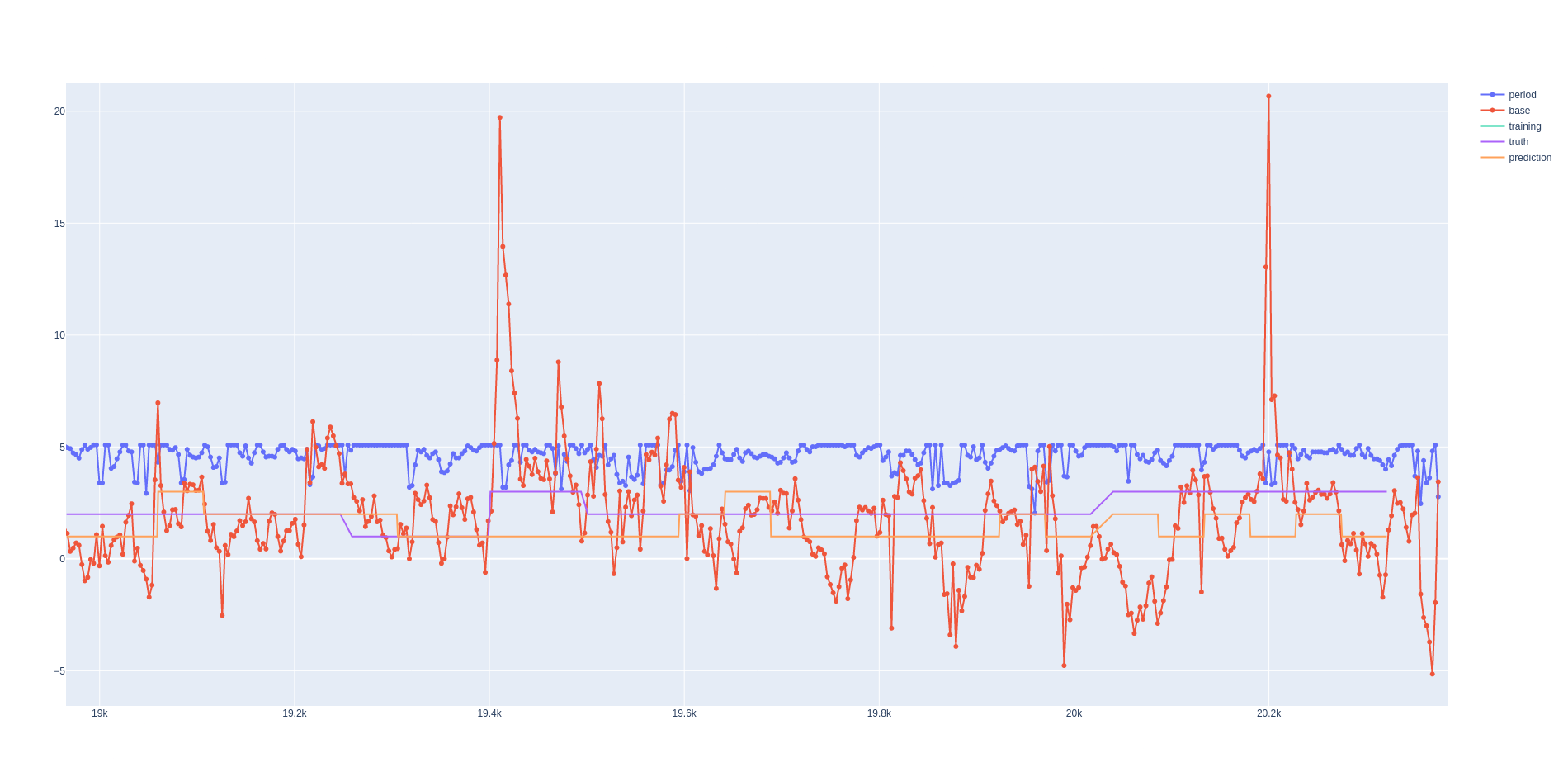}
	\caption{Attempt at classification. Excerpt from patient DB114-Fx1 represented by the model variables base and period in red and blue dotted lines, respectively. Classification categories range from 1 to 3. Predicted category in yellow, truth in purple.}
	\label{fig:ai-classifier}
	\vspace{-1em}
\end{figure}

In the first experiment, learning verification time, we first used our synthetic data. Then, we varied queries and, third, replaced the synthetic data by real data.
In all cases we applied a beam list to a single patient dataset, but we also tried to scale the number of beams from the typical 250 to 2500. Regarding the queries, we tried to vary between different kinds of queries, e.g., existential and universal, but did not notice significant differences.
The predictions were consistently at the lower end of verification time with more visible noise in the actual verification times. This can be seen visually in Figure~\ref{fig:ai-regressor}. Thus, this learning approach does not yield any opportunity for priorization.
Our explanation is that the synthetic data together with rather fast queries, that complete within a few hundred milliseconds, is rather unlearnable noise, because there is no discernable pattern within the parameters for longer durations. Also, simply increasing the number of queries from about 250 to about 2500 to have more data to learn may not help when the overall time becomes larger faster than any results are improving. Based on the experiments with real data, we could not make predictions on verification time that were significant enough to be of any help.

In the second experiment, we manually labeled a full motion dataset with three categories representing different levels of breathing periods, from calmer to more irregular. We applied a multi-layer perceptron classifier and varied the solver configuration with the options that \textit{scikit-learn} offers. As can be seen in Figure~\ref{fig:ai-classifier} almost all classifications are wrong, specifically when the period is spiking the level is classified as lower.
Based on the classification of a real breathing sequence, the prediction does not yield the desired results of differentiating more irregular from calmer breathing periods. Especially extreme movements are completely removed from the difficulty class. Thus, a better approach for classification is an algorithm specialised to detect more extreme cases in the movement data.
We refrained from integrating the results of our AI experiments in our beam scheduling application.

\section{Summary}\label{sec:summary}
Robotic radiation therapy with robotic ultrasound guidance represents an interesting area of application for OMC. We described and implemented an approach for beam scheduling which verifies feasibility of beam delivery during the treatment with OMC. Furthermore, we provide data and generated models. Our preliminary results show that, avoiding beam collision, reductions in the idle time ranging from \SIrange{16.02}{37.21}{\percent} are possible. Comparatively, a naive machine learning approach to predict beam delivery feasibility does not achieve the same performance as OMC while also being harder to reason about.

\FloatBarrier
\clearpage

\appendix
\section{Model declarations}\label{apx:declarations}
\subsection*{Global declarations}
\begin{lstlisting}[basicstyle=\ttfamily\footnotesize,numbers=none]
const double accuracy = 100.0;

const double period = 3469.0;
const double drift = 0.0;

double base = 2.5019;
double a[4] = { -0.1959, 0.0295, -0.0022, -0.0169 };
double b[4] = { -0.4023, 0.0294, 0.033, 0.013 };

double v[4];
double result;

double time;
broadcast chan step;

double frequency = 2 * 3.14159265358979323846 / period;
\end{lstlisting}

\subsection*{System declarations}
\begin{lstlisting}[basicstyle=\ttfamily\footnotesize,numbers=none]
Clock = Timer(38);

const double accrate = (100.0 - accuracy) / 15.0;

CMP = CoeffModifier(frequency, 1.0 * accrate * 0.0001, 10, 1000);
CMB = CoeffModifier(base, 1.0 * accrate * 0.25, 10, 1000);
TM1 = TermModifier(a[0], b[0], 1.0 * accrate * 0.1, 10, 1000);
TM2 = TermModifier(a[1], b[1], 1.0 * accrate * 0.1, 10, 1000);
TM3 = TermModifier(a[2], b[2], 1.0 * accrate * 0.1, 10, 1000);
TM4 = TermModifier(a[3], b[3], 1.0 * accrate * 0.1, 10, 1000);

First = FSTerm(v[0],a[0],b[0],frequency,1);
Second = FSTerm(v[1],a[1],b[1],frequency,2);
Third = FSTerm(v[2],a[2],b[2],frequency,3);
Fourth = FSTerm(v[3],a[3],b[3],frequency,4);

TermSummer = Summer(result,v,base);

system Clock, First, Second, Third, Fourth, TermSummer, CMP, CMB, TM1, TM2, TM3, TM4;
\end{lstlisting}

\subsection*{Automaton Parameters}
\begin{description}
	\item[FSTerm] \texttt{\footnotesize double \&v, double \&a, double \&b, double \&freq, int k}
	\item[Summer] \texttt{\footnotesize double \&s, double \&v[4], double \&offset}
	\item[Timer] \texttt{\footnotesize int rate}
	\item[TermModifier] \texttt{\footnotesize double \&a, double \&b, double rate, int minWait, int maxWait}
	\item[CoeffModifier] \texttt{\footnotesize double \&c, double rate, int minWait, int maxWait}
\end{description}

\subsection*{Automaton Declarations}
\begin{description}
	\item[Timer] \texttt{\footnotesize clock dt;}
	\item[TermModifier] \texttt{\footnotesize meta double m; clock t;}
	\item[CoeffModifier] \texttt{\footnotesize clock t;}
\end{description}

\section{Evaluation model}\label{apx:bbox}
\begin{figure}[hb]
	\centering
	\includegraphics[width=.9\textwidth,keepaspectratio]{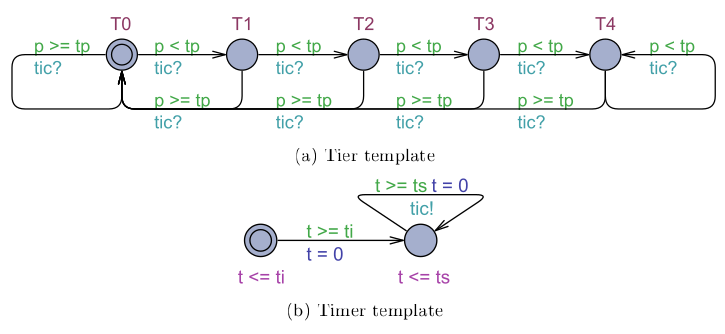}
	\caption{The Tier template is instantiated per data series, the Timer template is instantiated once. For each data series, $p$ is the current validity probability and $tp$ is the corresponding threshold. The probability is estimated by \uppaal{}-SMC that within an observed time period, observed values are predicted by the original model\cite{Rinast2015}.}
	\label{fig:evaluation-model}
\end{figure}

The evaluation model is directly used as per its introduction in Figure~7.6 of \cite{Rinast2015}, and shown here in Figure~\ref{fig:evaluation-model}.
It is queried by \texttt{E<> ((A1.T3 || A1.T4) \&\& (A2.T3 || A2.T4) \&\& (A3.T3 || A3.T4))} to check if a high enough error tier is reached through repeated condition violation. The probability \textit{p} that is referenced is obtained by querying the motion model using the query
\begin{align*}
	Pr[\diamond_{\leq t_o - t_m + t_+} t_o - t_- \leq t_p \leq t_o + t_+ \wedge x_0 - x_- \leq x_p \leq x_o + x_+],
\end{align*}
where $t_m$ is the model's creation time, $t_o > t_m$ the later observation time with $x_o$ the observed values, $t_+, t_-, x_+, x_-$ range parameters defining the bounding box around the observed $(t_o, x_o)$ and predicted $(t_p, x_p)$.

\FloatBarrier
\clearpage

\bibliographystyle{eptcs}
\bibliography{beamchecking}
\end{document}